\documentclass[pre,aps]{revtex4}
\usepackage{subfigure}
\usepackage{graphicx}
\def\strutdepth{\dp\strutbox}
\def\nw#1{\strut\vadjust{\kern-\strutdepth\vtop to0pt{\vss\hbox to\hsize
{\hskip\hsize\hskip5pt$\leftarrow$\hss\strut}}}{\em #1}}
\begin{document}

\title{Cusps in interfacial problems}
\author{J. Eggers$^1$ \& M. A. Fontelos$^2$}

\affiliation{
$^1$School of Mathematics,
University of Bristol, University Walk,
Bristol BS8 1TW, United Kingdom  \\
$^2$  Instituto de Ciencias Matem\'aticas,
 (ICMAT, CSIC-UAM-UCM-UC3M), \\ C/ Serrano 123, 28006
Madrid, Spain
        }
\begin{abstract}
A wide range of equations related to free surface motion in two 
dimensions exhibit the formation of cusp singularities either in time, 
or as function of a parameter. We review a number of specific examples, 
relating in particular to fluid flow and to wave motion, and show  
that they exhibit one of two types of singularity: cusp or swallowtail. 
This results in a universal scaling form of the singularity, and permits 
a tentative classification.
\end{abstract}
\maketitle

\section{Introduction}
The geometry of singularities of wavefronts, i.e. caustics, has
been much studied \cite{Nye99}. The caustics, themselves, exhibit
singularities such as cusps, which can be classified using
catastrophe theory \cite{Arnold84,PS78}. In the present review we
point out that a similar classification appears to apply in a variety
of nonlinear PDE problems, which describe the motion of a free
(fluid) surface.

The two-dimensional problems studied here can
be written as a mapping of the physical domain to the unit disc,
the free surface being represented as the circle. The appearance of a 
singularity is associated with the non-invertibility of this conformal map 
at a given value of time or of some other parameter. Since the problems 
studied here are endowed with a holomorphic structure, such non-invertibility 
has a very particular geometrical consequence: the formation of a cusp 
with a 2/3 exponent or of a swallowtail singularity. 
When looking at the mapping as it evolves in
time, or depends on a parameter, it is apparent that the mapping
remains smooth as the self-intersection occurs. For this reason
the curve can be written as an analytic function of a suitable parameter.
Performing a local expansion around the point of self-intersection,
the properties of the curve are determined by the lowest order
terms of the Taylor expansion in the parameter.

\begin{figure}
\begin{center}
\subfigure[]{
\includegraphics[scale=0.2]{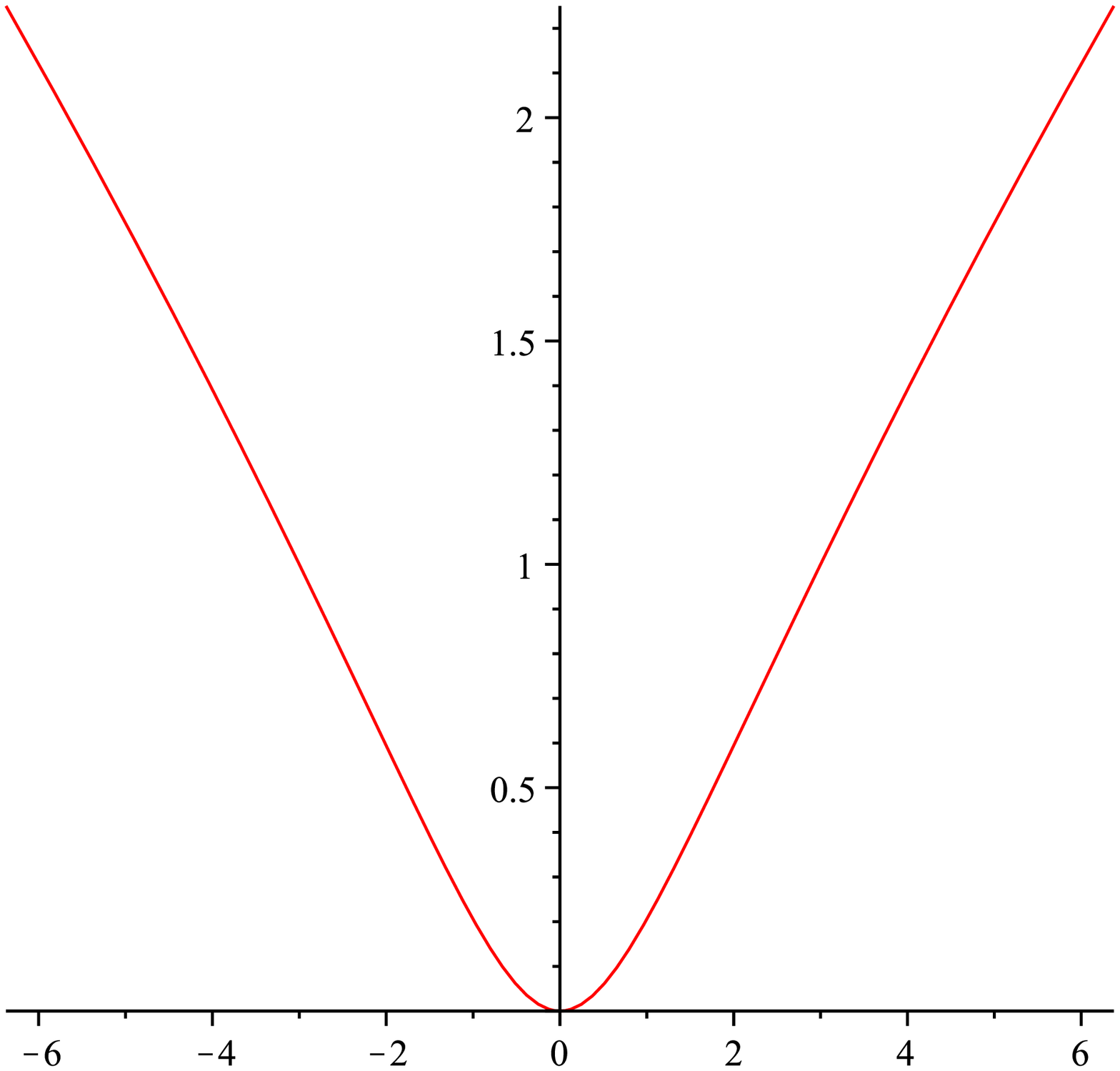}}
\subfigure[]{
\includegraphics[scale=0.2]{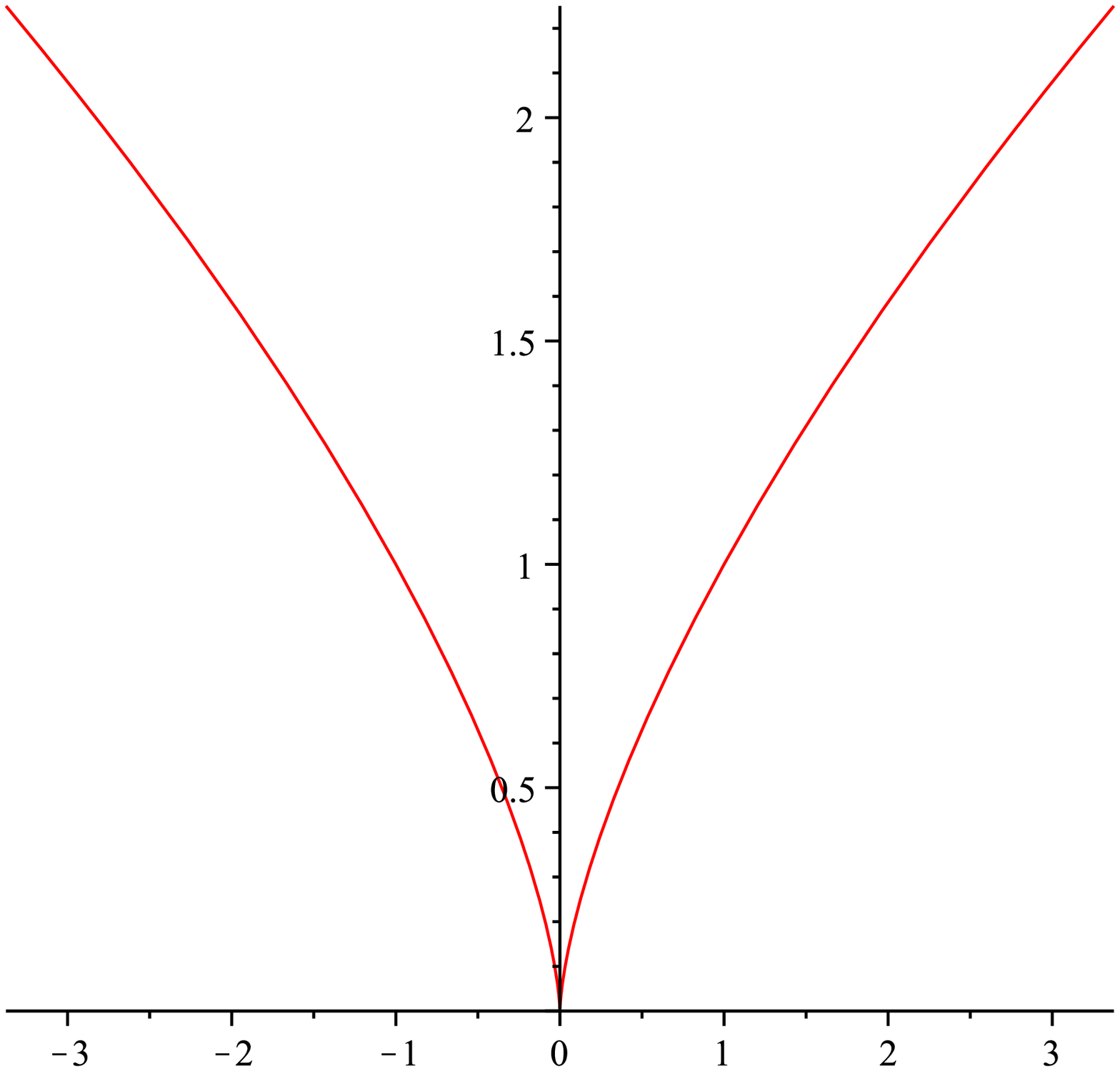}}
\subfigure[]{
\includegraphics[scale=0.2]{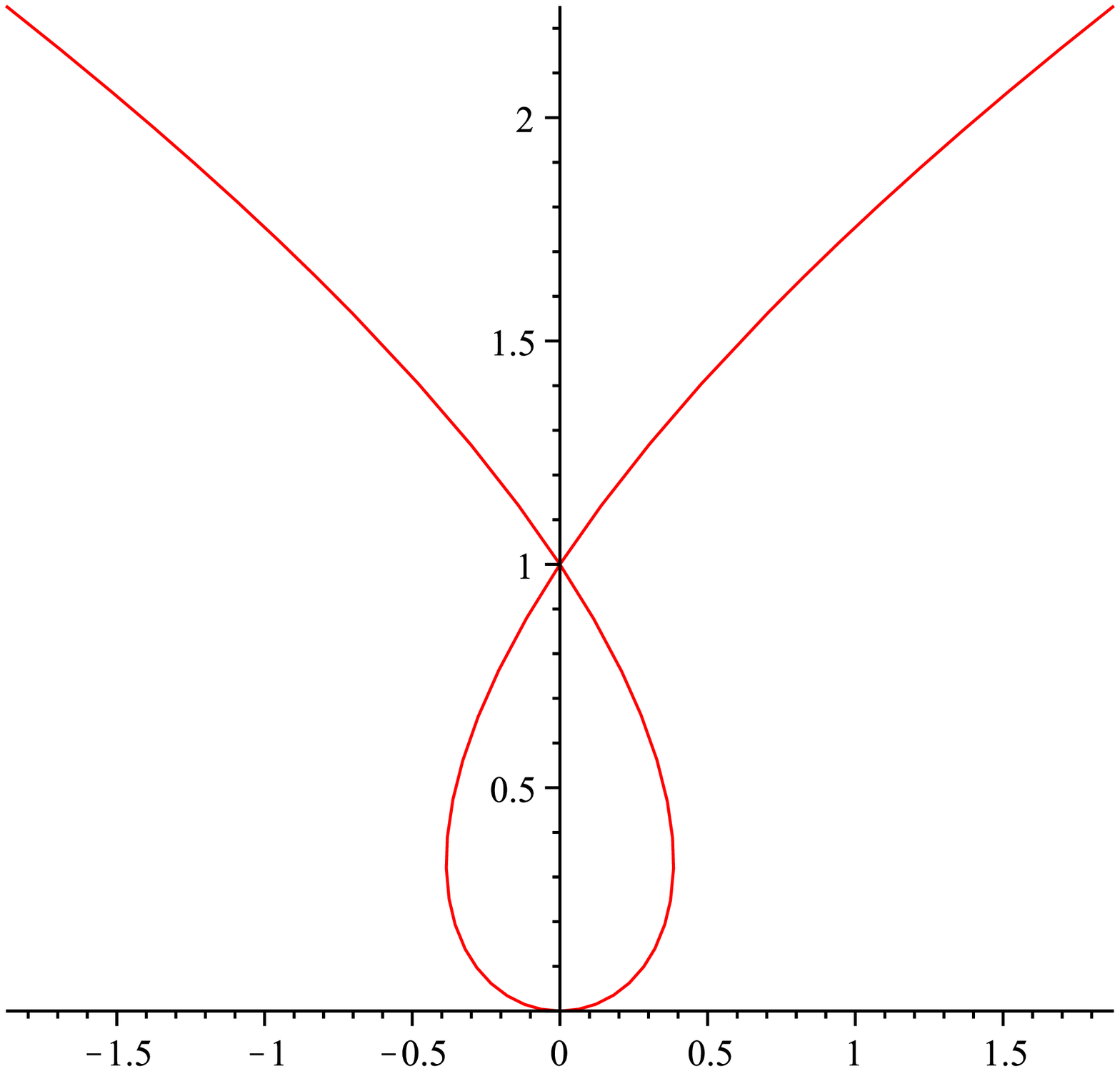}}
\caption{\label{cusp_fig}
The formation of a cusp, as described by (\ref{cusp}). Shown is
a non-intersecting curve ($\epsilon = 2$), a 2/3 cusp ($\epsilon = 0$),
and a loop ($\epsilon = -1$), from left to right. 
                   }
\end{center}
\end{figure}

The simplest case of self-intersection is that of the cusp.
Let us assume that the coordinates $x$ and $y$ can be expanded
into a Taylor series as function of some parameter, $\theta$,
and that we are interested in the neighborhood of $\theta=0$.
Since we describe phenomena up to arbitrary translations, the
generic description is $x=a_1 \theta$ and $y=a_2 \theta$.
However, by performing a rotation one can always make sure that
one of the coefficients ($a_2$, say), is zero. The singular
case corresponds to the situation where $a_1$ vanishes as well,
so we put $x=\epsilon \theta$, with the singularity at $\epsilon = 0$.

Of course we are now required to expand to higher order. The next
non-trivial order in $y$ gives $y = \theta^2/2$, where the coefficient
can be normalized by rescaling $\theta$. The expansion in $x$ has
to be pursued to third order, otherwise the curve would be degenerate
for $\epsilon = 0$. Thus we finally have:
\begin{equation}
x = \epsilon \theta + a \theta^3/3, \quad y = \theta^2/2,
\label{cusp}
\end{equation}
where $a$ is an arbitrary constant. The quadratic coefficient
in the expansion of $x$ can be eliminated by a shift in $\theta$,
with subsequent rotation and translation. Thus up to translations
and rotations, (\ref{cusp}) is the only generic way the self-intersection
of a curve in the plane can occur, as illustrated in Fig. \ref{cusp_fig}.
For $\epsilon < 0$ the curve self-intersects, at the critical
point $\epsilon = 0$ a cusp is formed. It is clear from (\ref{cusp})
that this cusp has the form $y = (x/a)^{2/3}$, i.e. it is associated
with a universal power law exponent. In catastrophe theory \cite{Arnold84},
(\ref{cusp}) is the curve associated with the ``cusp'' catastrophe.

\begin{figure}
\begin{center}
\subfigure[]{
\includegraphics[scale=0.2]{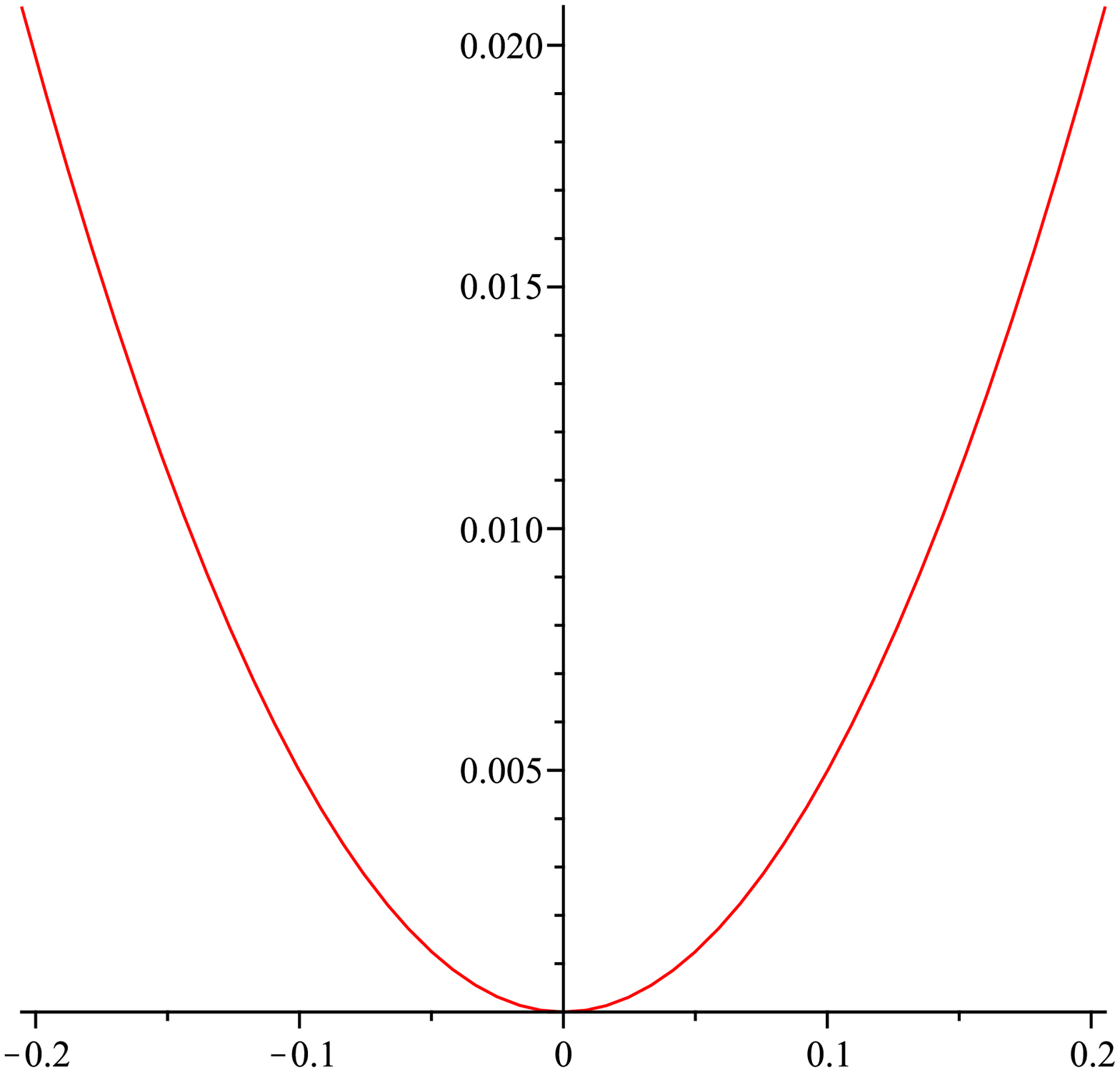}}
\subfigure[]{
\includegraphics[scale=0.2]{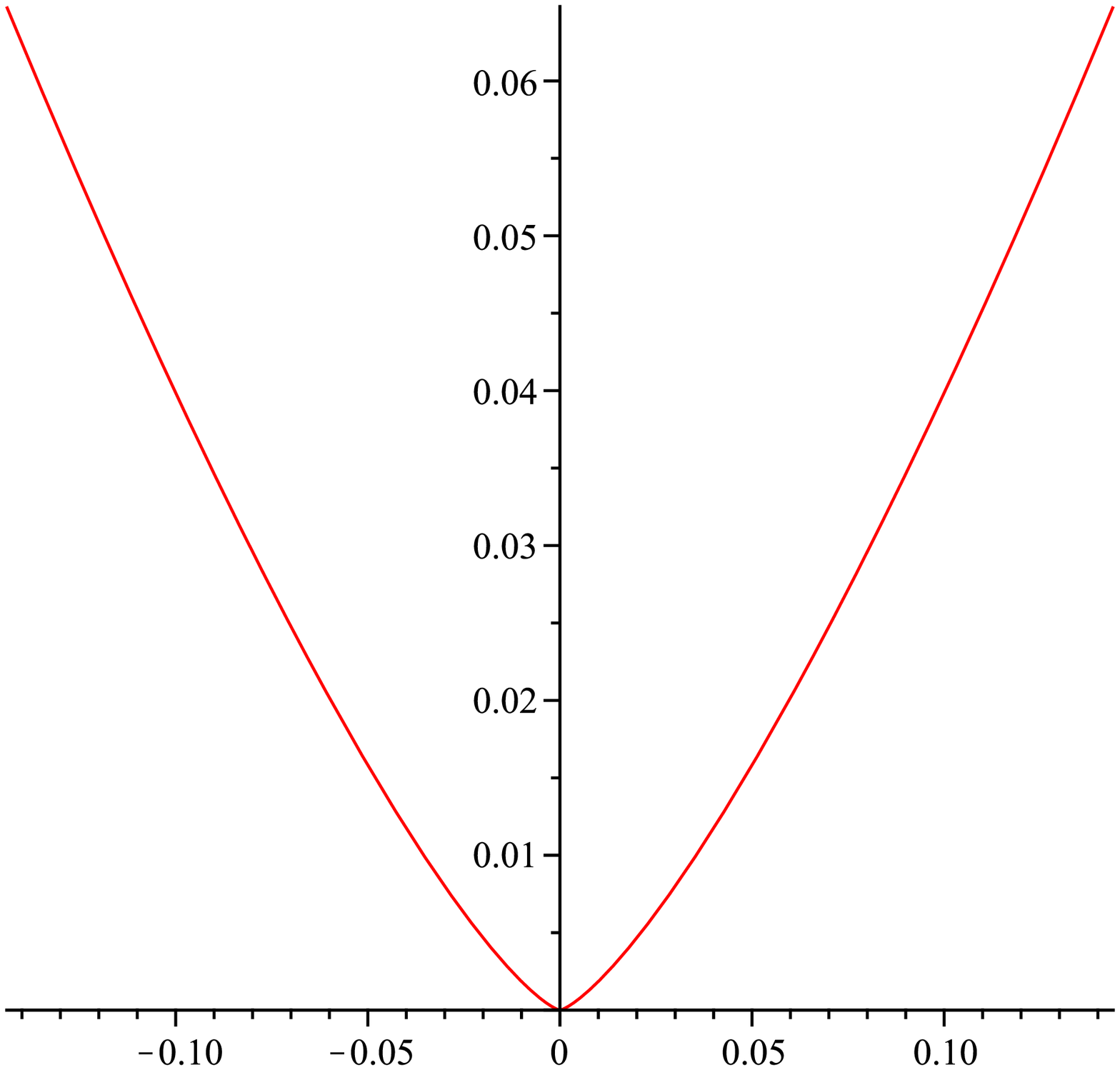}}
\subfigure[]{
\includegraphics[scale=0.2]{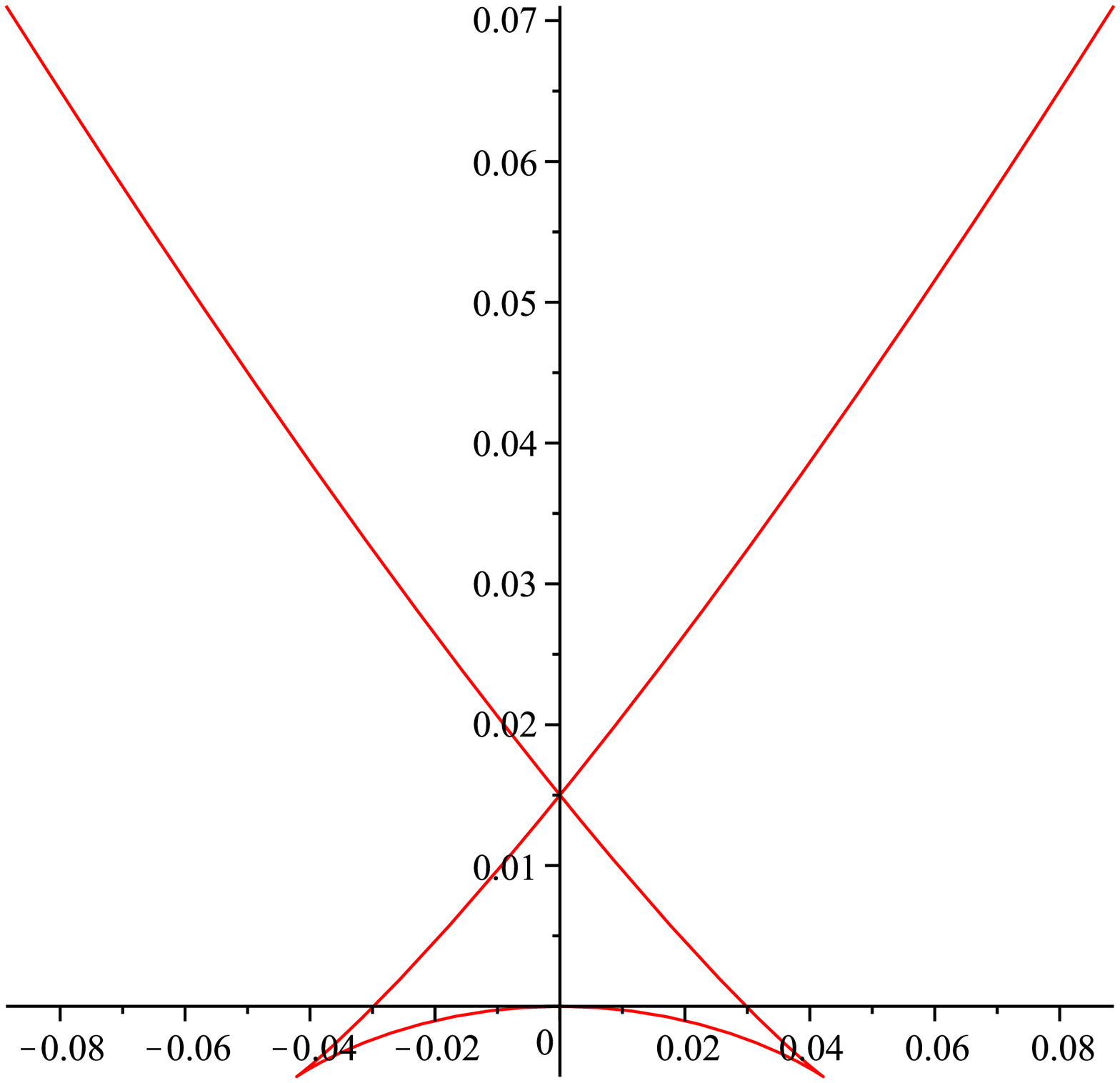}}
\caption{\label{swallow_fig}
The formation of a swallowtail, as described by (\ref{swallow}). Shown is
a smooth minimum, ($\epsilon = 1$, left), a minimum with a 4/3 singularity
($\epsilon = 0$, right), and a swallowtail or double cusp 
($\epsilon = -1/2$, right). }
\end{center}
\end{figure}

As we will see below, higher order singularities can also occur, which
presumably is due to a hidden symmetry of the problem. As a result, the
coefficient in front of the quadratic term vanishes in (\ref{cusp}):
$y = \epsilon_s\theta^2/2 + b_1\theta^3/3 + b_2\theta^4/4$. Once more,
the cubic term can be made to vanish by performing a rotation, so 
we arrive at the following generic form:
\begin{equation}
x = \epsilon_s \theta + a \theta^3/3, 
\quad y = \epsilon_s \theta^2/2 + b\theta^4/4, 
\label{swallow_gen}
\end{equation}
which exhibits a much milder singularity. Namely, for $\epsilon_s = 0$
the behavior is like $y \propto x^{4/3}$. For $\epsilon_s < 0$, the curve
splits into two cusp singularities of the form shown in 
Fig.~\ref{cusp_fig}, where the case $b>a$ corresponds to $\epsilon>0$
in the representation (\ref{cusp}), $b<a$ to $\epsilon<0$. 
In all the cases to be discussed below, we always find the particular 
case $a=b$, which means for $\epsilon_s<0$ the solution is exactly at the
cusp singularity. This is the swallowtail known from
catastrophe theory, which can be seen as a collision of two 
cusps (see Fig.~\ref{swallow_fig}). 
It has the universal form
\begin{equation}
x = \epsilon \theta + a \theta^3/3, \quad 
y = \epsilon \theta^2/2 + a\theta^4/4,
\label{swallow}
\end{equation}
once more with $\epsilon,a$ being parameters.

The two catastrophes (\ref{cusp}) and (\ref{swallow}) can occur
either as function of time or of some control parameter. Let us
introduce the time distance $t' = t_0-t$ to the singularity, and
assume the scaling $|\epsilon| = |t'|^{\gamma}$. The time before the
singularity $t'>0$ corresponds to $\epsilon>0$, and vice versa. 
Then 
\begin{eqnarray}
\label{cusp_sim}
x = |t'|^{3\gamma/2} X_c, \quad y = |t'|^{\gamma} Y_c, \\
X_c = \pm\sigma + a \sigma^3/3, \quad Y_c = \sigma^2/2,
\end{eqnarray}
is the self-similar form of the cusp singularity, and
\begin{eqnarray}
\label{swallow_sim}
x = |t'|^{3\gamma/2} X_s, \quad y = |t'|^{2\gamma} Y_s, \\
X_s = \pm\sigma + a \sigma^3/3, \quad
Y_s = \pm\sigma^2/2 + a\sigma^4/4
\end{eqnarray}
of the swallowtail singularity. In each case, the + or - signs correspond
to the similarity function before or after the singularity, respectively. 
We remark that the cusp singularity can be written as the following
cubic:
\begin{equation}
X_c^{2}=2Y_c(1\pm 2aY_c/3)^{2}.
\label{cubic}
\end{equation}

As we have mentioned before, the similarity function (\ref{swallow_sim}) 
valid {\it after} the singularity contains two cusp singularities, 
which occur for $\sigma = \pm 1/\sqrt{a}$. Thus if one writes 
$\sigma = \pm 1/\sqrt{a} + s$, shifts the cusp to the origin and performs 
a rotation, one obtains to leading order in $s$:
\begin{equation}
\left(\begin{array}{c}
          \pm a s^3\\
          (a+1) s^2
          \end{array}\right) = 
\left(\begin{array}{cc}
       1 & \mp\sqrt{a}\\
          \pm\sqrt{a} & 1
          \end{array}\right)
\left(\begin{array}{c}
X_s \pm 2/(3\sqrt{a}) \\
Y_s + 1/(4a)
       \end{array}\right)
\label{swallow_rot}
\end{equation}
The + or - signs correspond the left and right cusp, respectively. This 
demonstrates that $(X_s,Y_s)$ locally traces out a cusp after the swallowtail
singularity, as seen in Fig.~\ref{swallow_fig}. 

\begin{table}
  \begin{center}
    \leavevmode
\begin{tabular}{|l|l|l|l|}
\hline
Equation & Type & $\gamma$ & Section \\ \hline\hline
Wave fronts & swallowtail & 1 & \ref{sec:eik} \\ \hline
Hele-Shaw flow & cusp &  1/2 & \ref{sec:Hele} \\ \hline
Potential flow with free surface & swallowtail & // & \ref{sec:pot} \\ \hline
Porous medium equation & cusp & //  & \ref{sec:por} \\ \hline
Viscous flow with free surface & cusp & //  & \ref{sec:vis} \\ \hline
Born Infeld equation & swallowtail & 1  & \ref{sec:Born} \\ \hline
\end{tabular}
\end{center}
\caption{A summary of evolution equations discussed in this paper.
The classification as ``cusp'' or ``swallowtail'' refers to (\ref{cusp}) 
or (\ref{swallow}), respectively. In the case of time-dependent 
problems, the exponent $\gamma$ is defined by (\ref{cusp_sim}) or 
(\ref{swallow_sim}), depending on the type of singularity. 
     }
\label{list}
\end{table}
Table~\ref{list} summarizes the the problems to be studied in this 
paper, and cites the relevant sections. Each equation, to be discussed 
in more detail in the sections below, exhibits singularities which
can be classified as either being of the ``cusp'' or the ``swallowtail'' type.
Some are evolution equations, in which case we give the temporal 
scaling exponent $\gamma$, others exhibit singularities as function of 
a parameter. In each case, we attempt to give physically motivated examples. 

\section{The eikonal equation}
\label{sec:eik}
\subsection{Hamilton-Jacobi equation}
We consider the simplest case of wave propagation in a homogeneous 
medium, so the eikonal equation becomes $V_n=1$: the normal velocity 
of the wave front is constant everywhere. We will only consider 
wave propagation in two dimensions, so the wave front is a curve. 
A straightforward way to solve the eikonal equation is to 
consider $y=h(x,t)$ so that one can write the equation in the form
\begin{equation}
h_{t} =\sqrt{1+h_{x}^{2}}, \quad h(x,0)=h_{0}(x),
\label{eik_motion}
\end{equation}
which expresses the condition of constant normal velocity. 

To solve (\ref{eik_motion}), we pass to the ``particle'' description;
the equation for the wave front (\ref{eik_motion}) is the Hamilton-Jacobi 
equation for the Hamiltonian 
\begin{equation}
H(x,p)=-\sqrt{1+p^{2}},
\label{eik_hamilton}
\end{equation}
where $p = h_x$. The initial condition $(x_{0},h_{0}^{\prime }(x_{0}))$
yields a curve $(x,p)$ in phase space, and the trajectories will be
\begin{equation}
\frac{dx}{dt}=\frac{\partial H}{\partial p}=-\frac{p}{\sqrt{1+p^{2}}}, \quad
\frac{dp}{dt}=-\frac{\partial H}{\partial x}=0.
\label{eik_mech}
\end{equation}
Thus the solution to the particle problem is 
\begin{equation}
p = const=h_{0}^{\prime }(x_{0}), \quad
x = x_{0}-\frac{p}{\sqrt{1+p^{2}}}t=x_{0}-\frac{h_{0}^{\prime }(x_{0})}{
\sqrt{1+h_{0}^{\prime 2}(x_{0})}}t. 
\label{part_sol}
\end{equation}

We can obtain an explicit solution by noting that 
\[
\frac{dh}{dx} = \frac{dh}{dx_0}\frac{dx_0}{dx} = h_0'(x_0),
\]
where from (\ref{part_sol}) 
\[
\frac{dx}{dx_0} = 1 - \frac{h_0''}{\sqrt{1+h_{0}^3}} t. 
\]
Integrating the resulting expression for $dh/dx_0$, we finally obtain
\begin{equation}
x = x_0-\frac{h_0^{\prime}}{\sqrt{1+h_{0}^{\prime 2}}}t, \quad
h = h_0(x_0)+\frac{t}{\sqrt{1+h_{0}^{\prime 2}}}.
\label{explicit_sol}
\end{equation}
This is an explicit solution of (\ref{eik_motion}), which has the
additional advantage that it can be continued across any singularity
the wave front may encounter. 

A caustic is a place where rays meet, and thus $dx/dx_0=0$,
$dy/dx_0=0$. The two conditions turn out to be equivalent, and
one obtains
\begin{equation}
t_c = \frac{\sqrt{1+h_0'^2}^3}{h_0''} \equiv 1/\kappa.
\label{caus_cond}
\end{equation}
It is also clear that points on the caustic correspond to singularities
of the {\it wave front} \cite{Berry92}. The first singularity occurs
at a time $t_0$ corresponding to the maximum of the curvature, for which
there is optimal focusing. The universal structure is obtained by
expanding $h_0$ in a series, which without loss of generality only
contains even terms:
\begin{equation}
h_0 = a_1x_0^2 + a_2x_0^4 + \dots.
\label{h0_exp}
\end{equation}

\begin{figure}
\includegraphics[width=0.4\hsize]{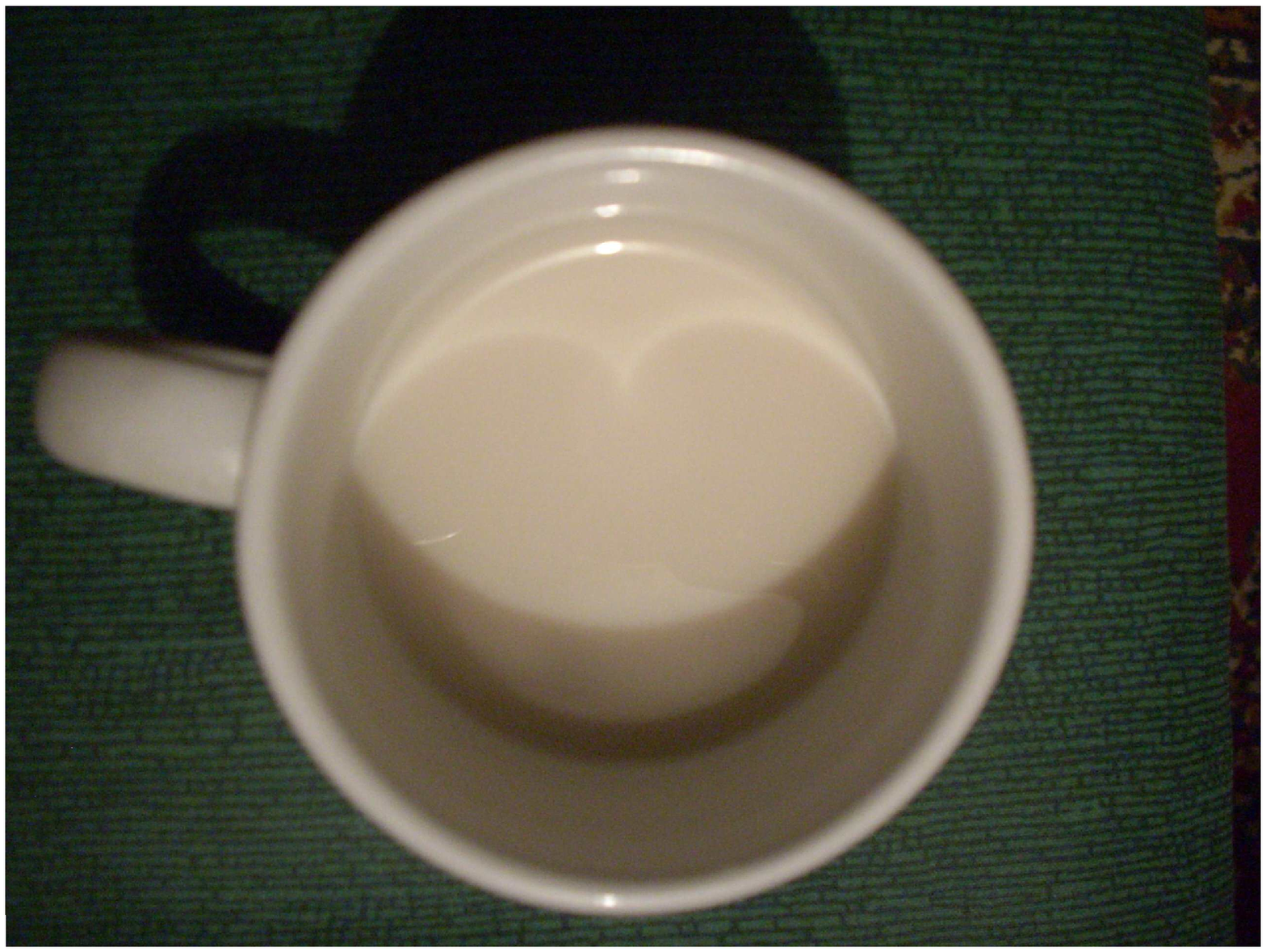}
\vspace{0.5cm}
\quad\quad
\includegraphics[width=0.35\hsize]{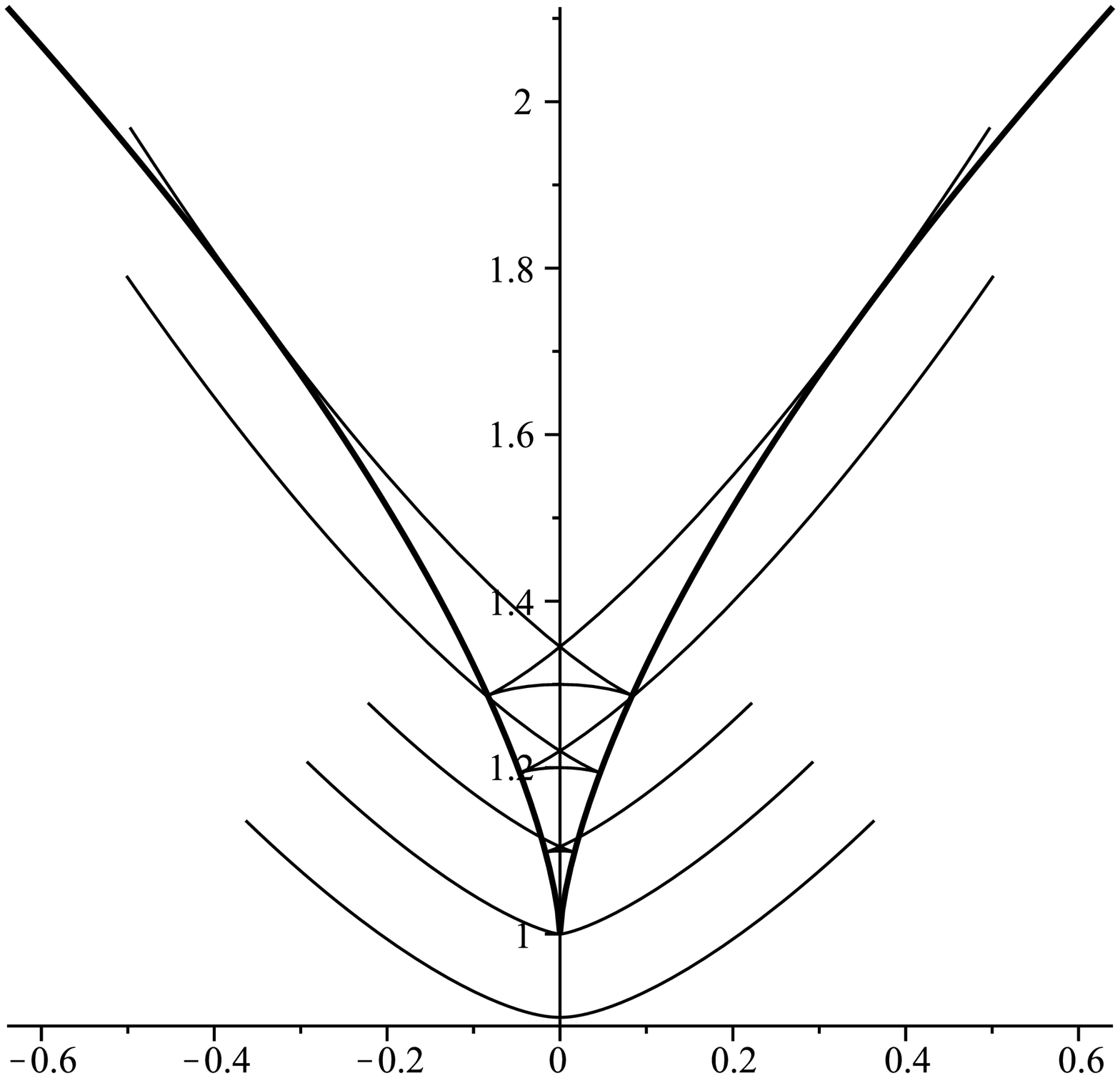}
\caption{
A caustic in a coffee cup. On the left, physical reality. 
On the right, the evolution of the wave front, as it passes 
through the swallowtail singularity. 
The caustic forms a 2/3-cusp,
which corresponds to a swallowtail in terms of the shape of
the wavefront.
    }
\label{coffee}
\end{figure}

The first singularity occurs for $t_0 = 1/(2a_1) +\dots$, and an
expansion leads precisely to (\ref{swallow_sim}), namely
\begin{eqnarray}
\label{eik_swallow}
&& x = t'\sigma + a \sigma^3/3 \\ \nonumber
&& y = 1/(2a_1)- t' + t'\sigma^2/2 + a \sigma^4/4,
\end{eqnarray}
where $a = 3(a_1^3-a_2)/(4a_1^4)$, and $\sigma = 2a_1x_0$;
the scaling exponent is $\gamma = 1$. After the singularity,
$t' < 0$, (\ref{eik_swallow}) is a swallowtail, and the two cusp 
points trace out the caustic. This corresponds to
the positions $\sigma_c = \pm\sqrt{-t'/a}$, and so to leading
order the equation of the caustic is
\begin{equation}
x_c = \pm\frac{-4t'^{3/2}}{a^{1/2}}, \quad
y_c = \frac{1}{2a_1} - t',
\label{caus_cusp}
\end{equation}
which is a normal 2/3-cusp, see Fig.~\ref{coffee}.

\subsection{conformal mapping}
There is a second way of solving the eikonal equation, closer to the methods
to be used in the next sections, based in the use of a complex
representation of the front. We present it here to stress the similarities
between the different problems studied in this paper. 
If one identifies the points of the curve 
$(x(\sigma ,t),y(\sigma ,t))$ as points $z(\sigma ,t)=x(\sigma ,t)+$ $%
iy(\sigma ,t)$ in the complex plane, and the velocity $%
\mathbf{u}=(u,v)$ as the complex number $u+iv$, then 
\begin{equation}
z_{t}=u+iv .
\label{complex}
\end{equation}
The tangent and normal vectors have a complex representation
\begin{equation}
\mathbf{t}=z_{s}, \quad \mathbf{n}=iz_{s}, 
\label{complex_tn}
\end{equation}
where $s$ is the arclength parameter.
Hence we can write
\[
\mathbf{u}\cdot \mathbf{n}=\mathrm{Re}\left( iz_{s}\overline{z_{t}}\right) =-
\mathrm{Im}\left( z_{s}\overline{z_{t}}\right),
\]
and the eikonal equation is simply
\begin{equation}
\mathrm{Im}\left( z_{s}\overline{z_{t}}\right) =-1, 
\label{eik}
\end{equation}
with the constraint $|z_s|=1$. Equivalently, one can introduce an arbitrary 
parameter $\sigma$ of the curve, and (\ref{eik}) becomes
or, equivalently,
\begin{equation}
\mathrm{Im}\left( z_{\sigma }\overline{z_{t}}\right) =-\frac{ds}{d\sigma},
\label{eik_equ}
\end{equation}
without the necessity of introducing a constraint. 

We now demonstrate that (\ref{eik_equ}) yields the same local solution 
as before, by verifying that (\ref{eik_swallow}) yields, at leading
order, a solution of (\ref{eik_equ}):
\begin{eqnarray*}
z_{\sigma } &=&(t^{\prime }+a\sigma ^{2})+i(t^{\prime }\sigma +a\sigma
^{3})\ ,\ z_{t}=-\sigma +i(1-\sigma ^{2}/2) \\
\mathrm{Im}\left( z_{\sigma }\overline{z_{t}}\right) &=&-(1+\frac{\sigma ^{2}%
}{2})(t^{\prime }+a\sigma ^{2}) \\
\frac{ds}{d\sigma } &=&\sqrt{(t^{\prime }+a\sigma ^{2})^{2}+\sigma
^{2}(t^{\prime }+a\sigma ^{2})^{2}}=(1+\frac{\sigma ^{2}}{2})(t^{\prime
}+a\sigma ^{2}+O(\sigma ^{4})) . 
\end{eqnarray*}

\section{Hele-Shaw flow}
\label{sec:Hele}
A Hele-Shaw cell consists of two closely spaced glass plates, partially
filled with a viscous fluid. The problem is to find the time evolution
of the free interface between fluid and gas. Here we consider the case 
that the fluid occupies a closed two-dimensional domain $\Omega$. 
Within $\Omega$, the pressure obeys $\Delta p=0$, with boundary conditions
\begin{eqnarray}
p &=&0  \label{c1} \\
V_n &=&-\nabla p\cdot \mathbf{n}  \label{c2}
\end{eqnarray}%
on the free surface $\partial \Omega$. We write $z=x+yi$ and
\[
p=\rm{Re}\Phi (z)
\]%
together with the conformal mapping
\begin{equation}
z=f(\xi,t), \quad \xi =re^{i\theta}, \label{c25}
\end{equation}
which maps $\left\vert \xi \right\vert =1$ onto $\partial \Omega $. 
If we consider
\[
\Phi (z(\xi ,t))=\log \xi \ \ \ \text{inside }\Omega, 
\]%
then condition (\ref{c1}) is automatically satisfied. Moreover
\[
-\nabla p\cdot \mathbf{n}=-\rm{Re}\left( \frac{\partial \Phi }{\partial z}%
iz_{s}\right) =-\rm{Re}\left( \frac{\partial \Phi }{\partial \xi }\frac{1}{%
z_{\xi }}iz_{s}\right) =-\rm{Re}\left( \frac{1}{\xi z_{\xi }}iz_{s}\right),
\]%
where $s$ is the arclength parameter. Notice that one can write for $\left\vert
\xi \right\vert =1$, using $\xi =e^{i\theta (s,t)}$
\[
z_{s}=z_{\xi }\xi _{s}=z_{\xi }\xi i\theta _{s},
\]%
and hence
\begin{equation}
-\nabla p\cdot \mathbf{n=}\theta _{s}.  \label{c3}
\end{equation}
Since 
\[
V_{n}=\rm{Re}\left( \overline{z_{t}}iz_{s}\right) =-\theta _{s}\rm{Re}%
\left( \overline{z_{t}}\xi z_{\xi }\right)
\]%
we arrive, combining this with (\ref{c3}), at the equation%
\begin{equation}
\rm{Re}\left( \overline{z_{t}}\xi z_{\xi }\right) =-1 . \label{c4}
\end{equation}
Noting that $\xi z_{\xi }=\frac{1}{i}z_{\theta}$, we finally obtain 
\begin{equation}
\rm{Im}\left( z_{\theta }\overline{z_{t}}\right) =-1. 
\label{c5}
\end{equation}
Equation (\ref{c5}) is identical to (\ref{eik}) except for the fact that 
the space variable is not the arclength parameter $s$, but the an 
arbitrary parameter $\theta$. The non-invertibility of $f(\xi,t)$ 
signals the appearance of a singularity. 

Now we will present an exact solution of (\ref{c4}), using a particular
(polynomial) ansatz for the mapping \ref{c25}
\cite{PK45a,PK45b,Ga45,How86}. We consider the simplest case of a 
quadratic:
\begin{equation}
f(\xi,t)=a_{1}(t)\xi +a_{2}(t)\xi^{2},
\label{Hele_map}
\end{equation}
and show that it leads to cusped solutions. However, we expect cusp 
formation to be a generic feature. The reason is
that the formation of a singularity is associated to the non-invertibility
of the conformal map $f(\xi ,t)$ and this is equivalent to $f^{\prime }(\xi
,t)$ being zero at some point $\xi _{0}$ (at the time $t_{0}$ of formation
of the singularity), which leads to a generic quadratic behavior of $f(\xi
,t_{0})=b_{0}+b_{2}(\xi -\xi _{0})^{2}$ near $\xi _{0}$. Other 
local expansions of $f(\xi ,t_{0})$ around  $\xi _{0}$, of the form $f(\xi
,t_{0})=b_{0}+b_{n}(\xi -\xi _{0})^{n}$ with $n>2$ are also possible. 
They would lead to different form of cusp, but cannot be generic 
since small perturbations of the initial data would produce quadratic terms. 

Inserting (\ref{c25}),(\ref{Hele_map}) 
into (\ref{c5}), we find a solution if the following system of 
ODEs is verified:
\begin{eqnarray}
&& a_{1}a_{1}^{\prime }+2a_{2}a_{2}^{\prime } = -1 \label{m1} \\
&& a_{1}a_{2}^{\prime }+2a_{2}a_{1}^{\prime } = 0. \label{m2} 
\end{eqnarray}%

Direct integration of the equations leads to
\begin{eqnarray}
&&\frac{1}{2}a_{1}^{2}+\frac{B^{2}}{a_{1}^{4}} =A-t \label{ms1} \\
&&A =\frac{1}{2}a_{1}^{2}(0)+a_{2}^{2}(0)\ ,\ B=a_{2}(0)a_{1}^{2}(0)
\label{ms2}
\end{eqnarray}%
A singularity occurs when $f$ fails to be invertible, that is%
\begin{equation}
\frac{df}{d\xi }=a_{1}+2a_{2}\xi =0\Rightarrow \xi =-\frac{a_{1}}{2a_{2}}=-1
\label{Hele_inv}
\end{equation}
or equivalently when
\begin{equation}
a_{1}-4\frac{B^{2}}{a_{1}^{5}}=0\Rightarrow a_{1}=(2B)^{\frac{1}{3}},a_{2}=%
\frac{1}{2}(2B)^{\frac{1}{3}} .
\label{Hele_inv2}
\end{equation}

From (\ref{ms1}) one can compute the singularity time
\[
t_{0}=A-\frac{3}{4}(2B)^{\frac{2}{3}}=\frac{1}{2}a_{1}^{2}(0)+a_{2}^{2}(0)-%
\frac{3}{4}(2a_{2}(0)a_{1}^{2}(0))^{\frac{2}{3}}. 
\]
Now let us consider a particular solution, for instance
\begin{equation}
a_{1}(0)=1\ ,\ a_{2}(0)=\frac{1}{16},
\label{Hele_part}
\end{equation}
so that 
\[
A=\frac{129}{256},B=\frac{1}{16},\ t_{0}=\left( \frac{3}{4}\right) ^{4}\ ,\
a_{1}(t_{0})=\frac{1}{2}\ ,a_{2}(t_{0})=\frac{1}{4}. 
\]%
This implies the formation of a singularity at $x(\pi,t_{0})=-\frac{1}{4}$
and $y(\pi,t_{0})=0$.

Local analysis leads to $a_{1}(t)=1/2+\widetilde{a}(t)$, where 
$3\widetilde{a}^{2}(t) \sim t_{0}-t$. Thus the scale factor 
$\widetilde{a}(t)$ is 
\begin{equation}
\widetilde{a}(t)\sim \frac{1}{\sqrt{3}}(t_{0}-t)^{\frac{1}{2}} \equiv
\frac{1}{\sqrt{3}}t'^{\frac{1}{2}}. 
\label{a_twiddle}
\end{equation}
and $a_{2}(t)=1/(16a_{1}^{2}(t)) \sim 1/4-\widetilde{a}(t)$. 
Writing $\xi =-1+\widetilde{\xi}$, we deduce
\begin{equation}
f(\xi ,t)=a_{1}(t)\xi +a_{2}(t)\xi ^{2}\sim -1-2\widetilde{a}(t)+3\widetilde{%
a}(t)\widetilde{\xi }+\frac{1}{4}\widetilde{\xi}^{2} .
\label{Hele_f}
\end{equation}

Since
\[
-1+\widetilde{\xi }=-e^{i(\theta -\pi )}=-1+(1-e^{i\widetilde{\theta }
})\simeq -1+\frac{\widetilde{\theta }^{2}}{2}-
\frac{\widetilde{\theta }^{4}}{4!}-i\widetilde{\theta }+
i\frac{\widetilde{\theta }^{3}}{3!}+O(\widetilde{\theta}^{5})
\]
one has
\[
f(\xi,t)\sim -1-2\widetilde{a}(t)+3\widetilde{a}(t)\left(
\frac{\widetilde{\theta}^{2}}{2}-\frac{\widetilde{\theta}^{4}}{4!} - 
i\widetilde{\theta }+i\frac{\widetilde{\theta}^{3}}{3!}\right) 
-\frac{1}{4}\widetilde{\theta}^{2}+\frac{1}{16}\widetilde{\theta}^4
- \frac{1}{4}\widetilde{\theta }^{3}i+O(\widetilde{\theta}^5),
\]
leading to the leading order contributions (all other terms are 
subdominant for small values of $\widetilde{a}(t)$ and 
$\widetilde{\theta})$: 
\begin{eqnarray}
x(\theta ,t)+1 &\sim &-2\widetilde{a}(t)-\frac{1}{4}\widetilde{\theta }^{2}
\label{Hele_sol1} \\
y(\theta ,t) &\sim &-3\widetilde{a}(t)\widetilde{\theta }-\frac{1}{4}
\widetilde{\theta }^{3}, 
\label{Hele_sol2}
\end{eqnarray}
which is the desired local solution we have been looking for. 

\begin{figure}
\includegraphics[width=0.5\hsize]{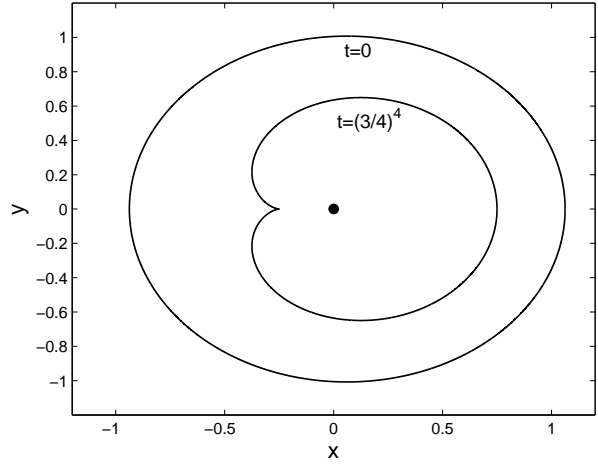}
\caption{
The formation of a cusp in Hele-Shaw problem with suction at the origin.
    }
\label{heleshaw}
\end{figure}

Defining
\begin{equation}
X_c=-\frac{y}{6t'^{3/4}},\ Y_c= -\frac{x+1}{6t'^{1/2}} - \frac{1}{3\sqrt{3}},\
\Theta =\frac{\widetilde{\theta}}{2\sqrt{3}t'^{1/4}},
\label{Hele_def}
\end{equation}
one obtains the cusp (\ref{cusp_sim}), with $a = 9\sqrt{3}$ and 
the scaling exponent $\gamma = 1/2$. A different choice of initial 
conditions (\ref{Hele_part}) will of course lead to a different value
of the parameter $a$. In Figure \ref{heleshaw} we represent the 
interface profiles corresponding to the example developed above at the 
initial time and at the time of formation of the cusp.

\section{Potential flow}
\label{sec:pot}
We consider the two-dimensional flow of an ideal fluid below a free surface.
Inside the fluid, the fluid velocity ${\bf u} = (u,v)$ satisfies
\begin{equation}
{\bf u} = \nabla\phi, \quad \triangle \phi = 0,
\label{pot}
\end{equation}
where $\phi$ is the velocity potential. The free surface is convected
by the fluid velocity, and the free surface is at constant pressure.
We consider the simplest case of steady flow, as well as no body or
surface tension forces. According to Bernoulli's equation, the fluid
speed then has to be constant on the free surface.

Exact solutions to the flow problem can be found if the fluid
domain is bounded by free surfaces and straight solid boundaries
alone \cite{Batch67}, by mapping the fluid domain onto the upper 
half of the complex plane, which we denote by $\zeta$. Here, we consider only
the even simpler case of only free boundaries, i.e. that of a two-dimensional
fluid drop. To this end, one introduces the complex potential 
$w = \phi + i\psi$, where $\psi$ is the stream function. 
The derivative of $w$ gives the fluid velocity:
\begin{equation}
\frac{dw}{dz} = u - iv \equiv q e^{-i\theta},
\label{compl_pot}
\end{equation}
where $q$ is the particle speed. 

On the fluid 
boundary, $\psi$ is constant, and this constant can be chosen to vanish.
Thus $w(\zeta)$ is real on the real axis (the boundary of the fluid 
drop), and so $dw/d\zeta$ must be real as well. 

Following Hopkinson \cite{Ho98} we drive a fluid motion inside the 
drop by placing singularities inside the drop. We will consider 
the case of a vortex dipole and a vortex at the same 
point. We have 
\begin{equation}
\frac{dw}{d\zeta} = \frac{dw}{dz}\frac{dz}{d\zeta}, 
\label{trans_w}
\end{equation}
where $dz/d\zeta$ contains no singularities in the upper $\zeta$ plane,
since the representation of the fluid flow must be conformal. But this
means that $dw/d\zeta$ must have the same singularities as $dw/dz$, 
except that we now have the freedom to choose the position of the two 
singularities and the orientation of the vortex doublet arbitrarily. 
Thus if we choose the two singularities at $\zeta = i$, and the 
orientation of the doublet toward the positive real axis, the potential 
for the doublet must locally look like 
$w\propto 1/(\zeta-i) -im\ln(\zeta-i)$. Here $m$ is the relative 
strength of the vortex. 

To insure that $dw/d\zeta$ also obeys the boundary condition, one has
to add image singularities at $\zeta = -i$:
\begin{equation}
\frac{dw}{d\zeta} = \frac{1}{(\zeta-i)^2} + \frac{1}{(\zeta+i)^2}
-im\left(\frac{1}{\zeta-i} - \frac{1}{\zeta+i}\right). 
\label{image_w}
\end{equation}
Depending on the value of $m$, two different cases arise. For 
simplicity, we only consider the case $m<1$, in which 
\begin{equation}
\frac{dw}{d\zeta} = \frac{2(1+m)(\zeta^2-\gamma^2)}{(\zeta^2+1)^2},
\label{dwdzeta}
\end{equation}
Where $\gamma = \sqrt{(1-m)/(1+m)}$ real. 

Manifestly, $dw/d\zeta$ is real on the real axis, and has the 
right singularities at $\zeta =i$. However, the information 
contained in (\ref{image_w}) is not enough to reconstruct the 
mapping $z(\zeta)$ we are after. Following Kirchhoff \cite{Kirch69} 
and Planck \cite{Planck}, we define another function 
\begin{equation}
\Omega = \ln\frac{dz}{dw} \equiv -\ln q + i\theta,
\label{omega}
\end{equation}
which will also be represented in the $\zeta$-plane. 
Since $q$ is constant along the free surface, and choosing 
units such that $q=1$, the function $\Omega$ will be purely 
imaginary on the real $\zeta$ axis. To find $\Omega$, we once
more proceed in two steps. 

First, we find the singularities of $\Omega$. From the definition,
\begin{equation}
\Omega = -\ln\frac{dw}{d\zeta} + \ln\frac{dz}{d\zeta},
\label{omega_zeta}
\end{equation}
where the second contribution is conformal in the upper half
plane. This means $\Omega$ has singularities only for $\zeta=i$,
where it behaves like $\Omega \propto 2\ln(u-i)$. Second,
we have to make sure that $\Omega$ is imaginary for real $\zeta$,
which is achieved by 
\begin{equation}
\Omega = 2\ln\frac{\zeta-i}{\zeta+i}.
\label{omega_final}
\end{equation}
Now we use the fact that 
\begin{equation}
\frac{dz}{d\zeta} = \frac{dz}{dw}\frac{dw}{d\zeta} = 
e^{\Omega}\frac{dw}{d\zeta} = \frac{2(1+m)(\zeta^2-\gamma^2)}{(\zeta+i)^4}.
\label{omega_int}
\end{equation}

\begin{figure}
\includegraphics[width=0.45\hsize]{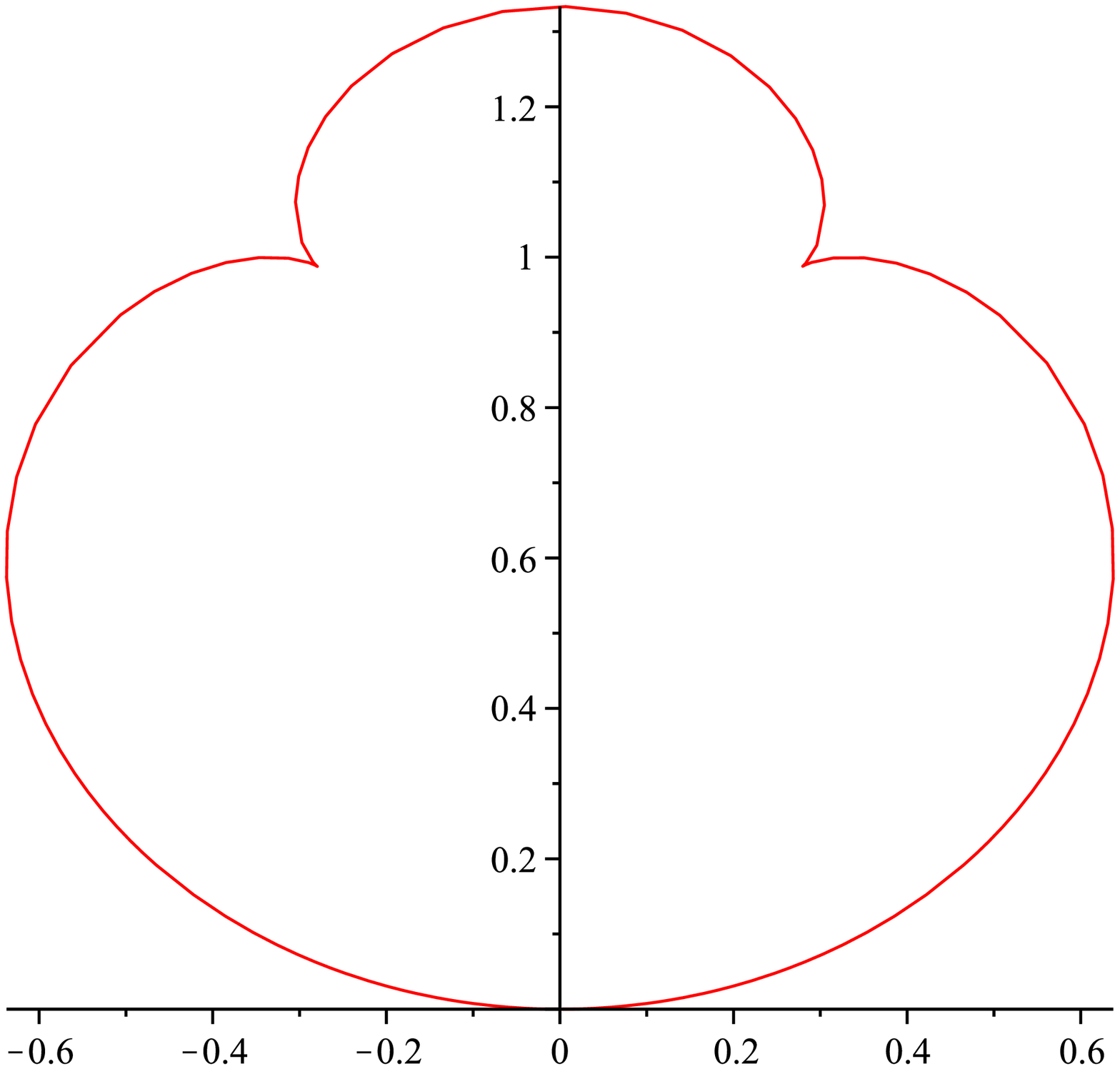}
\includegraphics[width=0.45\hsize]{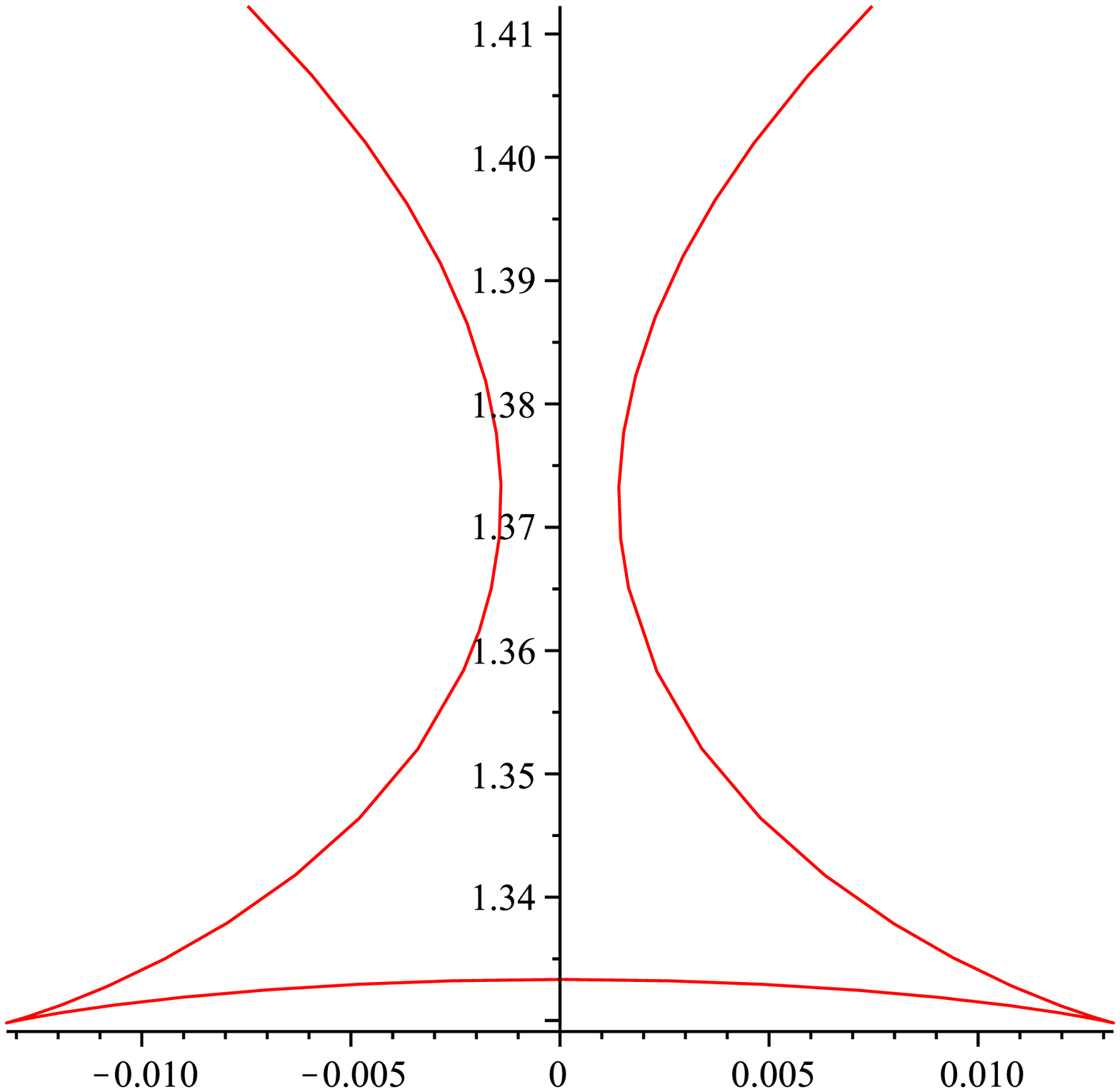}
\caption{
On the left, we show the entire drop for $m=1/3$. 
On the right, a closeup of upper part of the drop for $m=0.94$. 
The swallowtail has opened, and there is no more self-intersection. 
    }
\label{cusp_fig_ent}
\end{figure}

This expression can be integrated to find the transformation $z(\zeta)$
between the fluid domain and the $\zeta$-plane:
\begin{equation}
\frac{z}{2(1+m)} = -\frac{1}{\zeta+i} + \frac{i}{(\zeta+i)^2}
+ \frac{1+\gamma^2}{3(\zeta+i)^3}. 
\label{z_zeta_int}
\end{equation}
From the real and imaginary part of this expression, we obtain 
\begin{eqnarray}
&& x=-(2(1+m))\zeta\frac{3\zeta^4-(1+\gamma^2)\zeta^2+3\gamma^2}
{3(\zeta^2+1)^3} \\
&& y=(2(1+m))\frac{6\zeta^4+3(1-\gamma^2)\zeta^2+1+\gamma^2}
{3(\zeta^2+1)^3}.
\label{drop_shape2}
\end{eqnarray}
The mapping is such that $\zeta=\pm\infty$ gets mapped to the origin. 
A typical drop shape, for $m=1/3$, is shown in Fig.~\ref{cusp_fig_ent} 
(left); it exhibits two cusps. This feature is generic, in that it 
exists for a continuous range of values $0.9427\ge m \ge 0$. 
On the right of show a closeup of the top of the drop, close to the 
upper end of his range. 

\begin{figure}
\includegraphics[width=0.45\hsize]{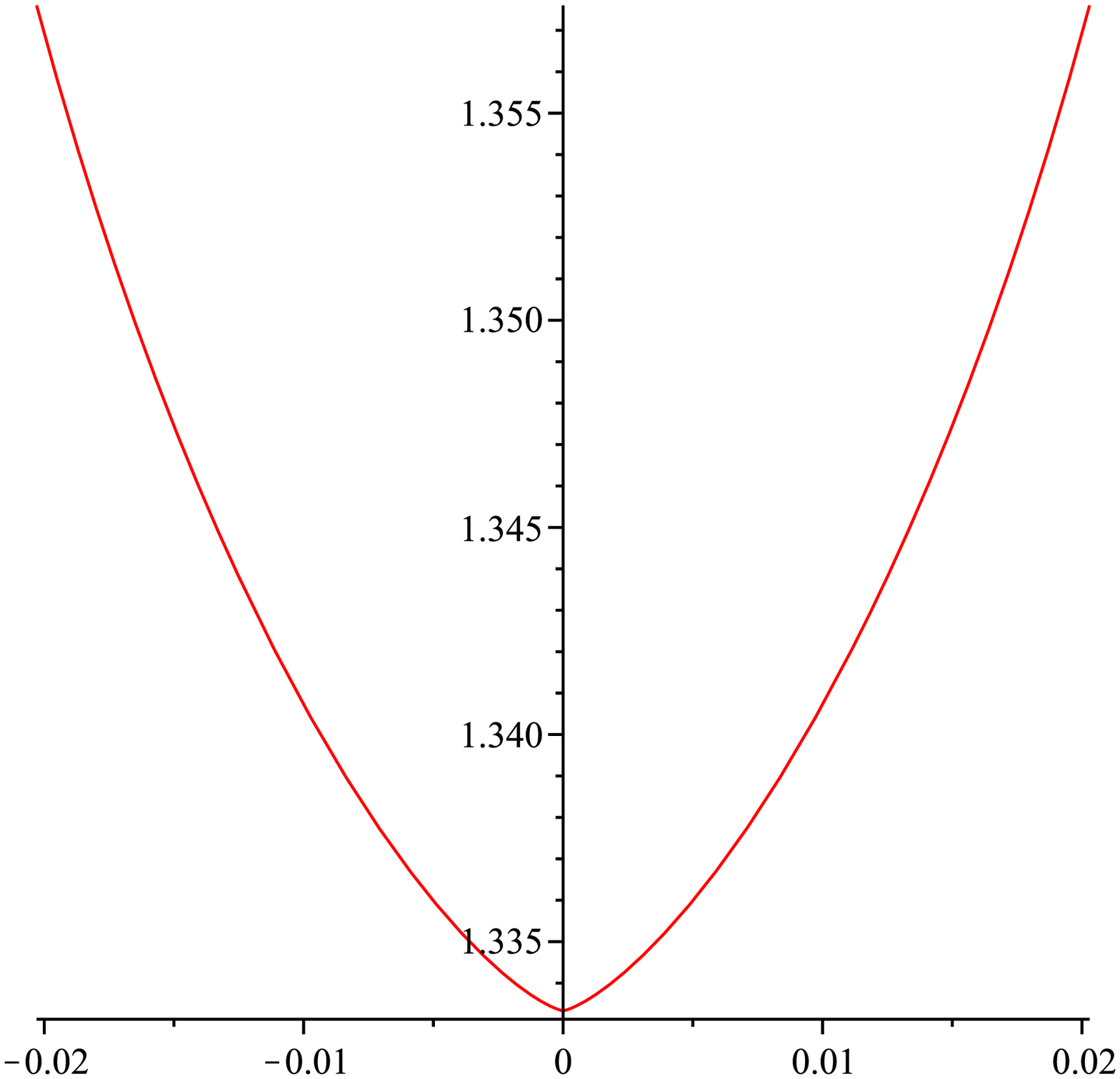}
\includegraphics[width=0.45\hsize]{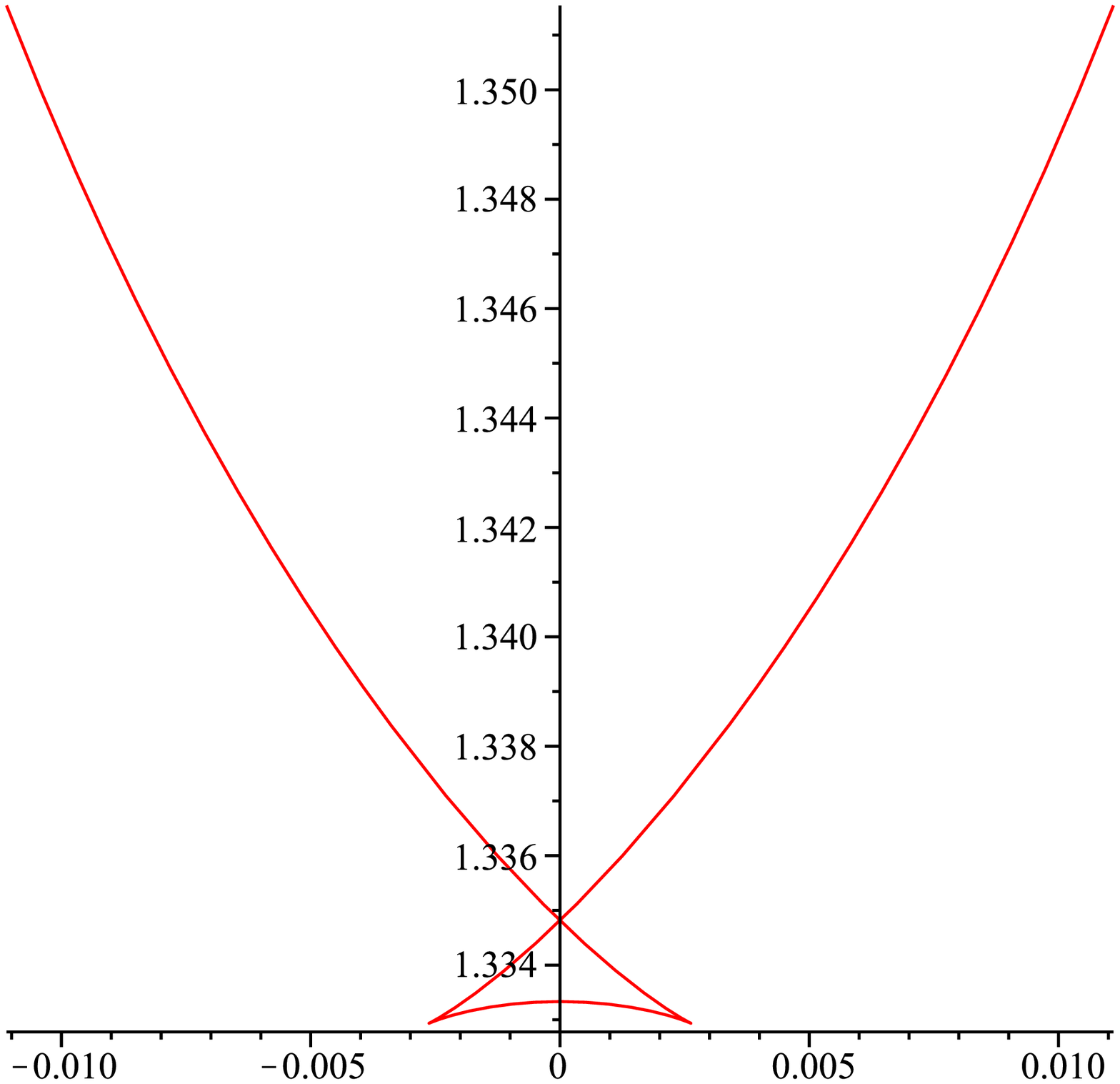}
\caption{
The birth of the swallowtail, as described by (\ref{drop_shape2}). 
For $m=1$ (left) one is exactly at the transition, for $m=0.98$ (right)
one finds a swallowtail. 
    }
\label{cusp_fig_loc}
\end{figure}

The origin of the double cusp lies in a swallowtail transition 
for $m\approx 1$ that occurs at the point $x = 0$, $y = 4/3$ in 
real space. Namely, a local expansion of (\ref{drop_shape2})
gives with $m = 1 + \epsilon$
\begin{eqnarray}
&& x=2\epsilon \zeta + 4\zeta^3/3 \\
&& y=4\epsilon\zeta^2 + 4\zeta^4,
\label{hopkinson}
\end{eqnarray}
where $\zeta$ is taken as real. The same expression results if the 
shape of the drop valid for $m >1$ is expanded locally. 
The corresponding swallowtail is shown in Fig.~\ref{cusp_fig_loc} 
using the full solution Fig.~\ref{drop_shape2}. For $m$ very close 
to the transition, the free surface must self-intersect, as shown in
Fig.~\ref{cusp_fig_loc} (right). However, below a value of about
$m \approx 0.9427$ the self-intersection disappears. 


Another exact solution of potential flow that exhibits
a cusp, but {\it in the presence of gravity} was found
by Craya and Sautreaux \cite{Cr49,Sau01,DH95}; liquid is
layered above a two-dimensional ridge.
At the crest of the ridge, which has opening angle $2\pi\gamma$,
there is a sink. For the special case $\gamma = 1/3$ there is an exact
solution, given by
\begin{equation}
\frac{dz}{dl} = -i(2/3)^{1/3}\frac{1-l}{l^{1/3}(1+l)^{4/3}},
\label{SC}
\end{equation}
where $l = \exp(-i\theta)$, $-\pi\le\theta\le\pi$.
From a local analysis we find that
\begin{equation}
x = -\frac{3^{2/3}}{36}\theta^3, \quad y = -\frac{3^{2/3}}{12}\theta^2,
\label{SC_loc}
\end{equation}
which is once more a 2/3 cusp. Of course this is to be expected,
since gravity cannot change the local behavior near a cusp.
However it is not clear where this comes from in terms of the
swallowtail described above.

\section{Porous medium equation}
\label{sec:por}
Another problem, closely related to the above, concerns the two-dimensional
flow of oil in a porous medium. The oil is layered above heavier water,
and is withdrawn through a sinkhole. The interface between the oil and
the water is deformed, and forms a cusp at a critical flow rate.
Inside the oil domain, one has to solve Laplace's equation for the
velocity potential, cf. (\ref{pot}). In the stationary case, one
has the usual condition of vanishing normal velocity on the
free surface. However, from a pressure balance which includes the hydrostatic
pressure, one obtains \cite{ZH96,ZHB97}:
\begin{equation}
u^2 + v^2 = Kv,
\label{por_bound}
\end{equation}
where $u,v$ are the horizontal and vertical components of the
velocity, respectively.

Once more hodograph methods can be applied, but the problem can be
solved only in the critical case at which there is cusp. Namely, in the 
subcritical case where there interface is smooth, $u=v=0$ at the stagnation 
point below the sink. In the presence of a singularity, $u=0, v=K$ at
the singularity, and the free surface gets mapped onto a circular arc
in the hodograph plane. The resulting interface shape is
\begin{eqnarray}
&& x = -\frac{2}{c\pi^2}\int_{-\beta}^{\sigma}\left(
\frac{\sqrt{a+1}}{\sigma_l}+
{\rm arctanh}\sqrt{a+1}\right)
\left(\frac{1}{a-\sigma_l}-\frac{1}{a-\sigma_s}\right) \label{por_shape1}\\
&& y=\frac{1}{c\pi}\left(\ln\frac{\sigma_l-\sigma}{\sigma-\sigma_s}-
\ln\frac{\sigma_l-1+\beta}{1-\beta-\sigma_s}\right),
\label{por_shape2}
\end{eqnarray}
where $\sigma$ parameterizes the curve and $\sigma_l$ and $\sigma_s$ 
are determined from implicit equations involving $c$ and $\beta$. 

For $\beta=1$ the curve defined by (\ref{por_shape1}),(\ref{por_shape2})
develops into a cusp. A local expansion yields
\begin{equation}
x = A\int_{-1}^{\sigma}\sqrt{1+a} da = \frac{2A}{3}(\sigma+1)^{3/2},
\label{por_exp}
\end{equation}
where $A$ is a constant. The $y$-coordinate is linear in $t+1$,
thus (\ref{por_exp}) describes the usual 2/3 cusp. For $\beta < 1$ the
interface self-intersects, at the line of symmetry, so this appears
to be the generic cusp scenario. Numerical results confirm  that the 
curve is smooth before the cusp forms (subcritical case), and a cusp 
forms in agreement with the exact solution 
(\ref{por_shape1}),(\ref{por_shape2}). 

\section{Viscous flow}
\label{sec:vis}
Here the flow is governed by the Stokes equation, which in
two dimensions can be written in terms of the stream function
$\psi$, with $u = \psi_y$ and $v = -\psi_x$. The stream function
obeys the biharmonic equation $\triangle^2\psi = 0$. In a stationary
state, which we are considering, the surface of the fluid is a line
with $\psi = const$. On this surface, we also have the surface stress
condition
\begin{equation}
\sigma_{ij}n_j = \gamma\kappa n_i,
\label{stress}
\end{equation}
where $\gamma$ is surface tension and $\kappa$ the curvature of
the interface. Once more the flow is driven by singularities, such as a vortex
dipole \cite{JM92}.

A complex formulation of this problem was developed by
Richardson \cite{R68}. The stream function is written as
\begin{equation}
\psi = Im(f(z) + \overline{z}g(z)),
\label{stream}
\end{equation}
where $f$ and $g$ are analytic functions. The boundary conditions
at the free surface can be shown to be 
\begin{equation}
Im\left[\left(\overline{\frac{dz}{ds}}\right)f(z) +
\overline{z}g(z)\right] = \frac{\gamma}{4\eta},
\quad f(z) + \overline{z}g(z) = 0,
\label{psi_bound}
\end{equation}
where $\eta$ is the viscosity of the fluid. 

In \cite{JM92}, the complex formulation (\ref{psi_bound}) was used 
to calculate the following model problem: a vortex dipole of 
strength $\alpha$ is located at a distance $d$ below a free surface
of infinite extend. Surface tension is included in the description,
but the effect of gravity is neglected. The deformation of the free 
surface by the viscous flow is determined by the capillary number 
\begin{equation}
Ca = \frac{\alpha\eta}{d^2\gamma},
\label{Ca_def}
\end{equation}
which measures the ratio of viscous forces over surface tension 
forces. The solution of the problem is too involved to be 
presented here. The exact surface shape, in units of $d$, is given 
by the function
\begin{eqnarray}
\label{surf_JM}
&& x = a\cos\theta + (a+1)\frac{\cos\theta}{1+\sin\theta}, \\
&& y = a(1+\sin\theta).
\end{eqnarray}

The parameter $a$ is determined from the equation 
\begin{equation}
4\pi Ca = \frac{-a(3a+2)^2K(m)}{1+a+\sqrt{-2a(a+1)}},
\label{a_equ}
\end{equation}
where 
\begin{equation}
m = \frac{2}{(-2a/(a+1))^{1/4}+((a+1)/(-2a))^{1/4}}
\label{m_equ}
\end{equation}
and $K$ is the complete elliptic integral of the first kind:
\begin{equation}
K(m) = \int_0^{\pi/2}\frac{d\theta}{\sqrt{1-m^2\sin^2\theta}}. 
\label{K_comp}
\end{equation}
In (\ref{a_equ}) we have only reported the form of the equation 
for the more relevant case $Ca>0$. Asymptotic analysis of 
(\ref{a_equ})-(\ref{K_comp}) reveals that for large $Ca$, 
\begin{equation}
a = -\frac{1}{3} + \epsilon, \quad \epsilon\approx 
\frac{32}{9} \exp\left\{-16\pi Ca\right\}. 
\label{a_asymp}
\end{equation}

\begin{figure}
\includegraphics[width=0.45\hsize]{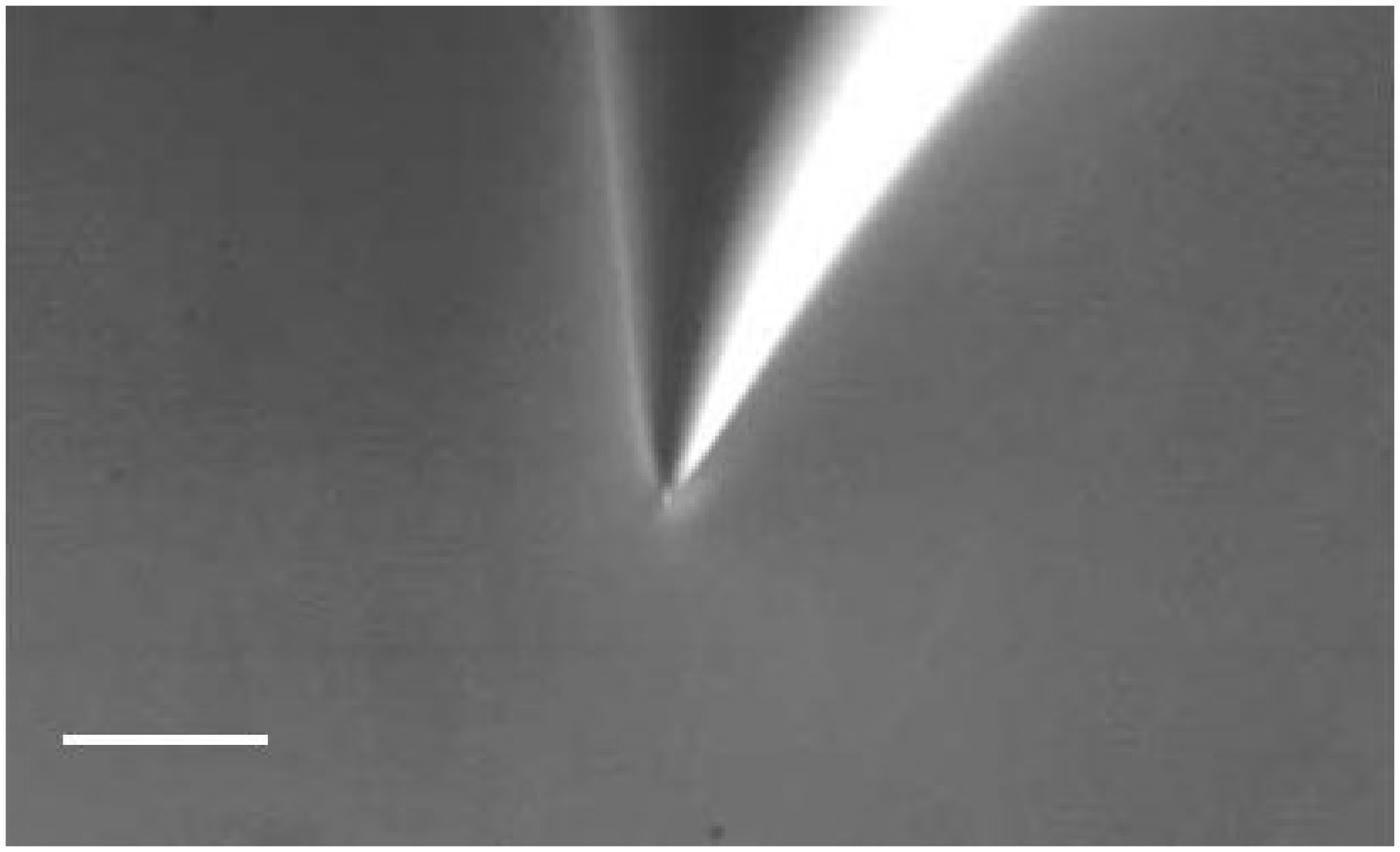}
\includegraphics[width=0.45\hsize]{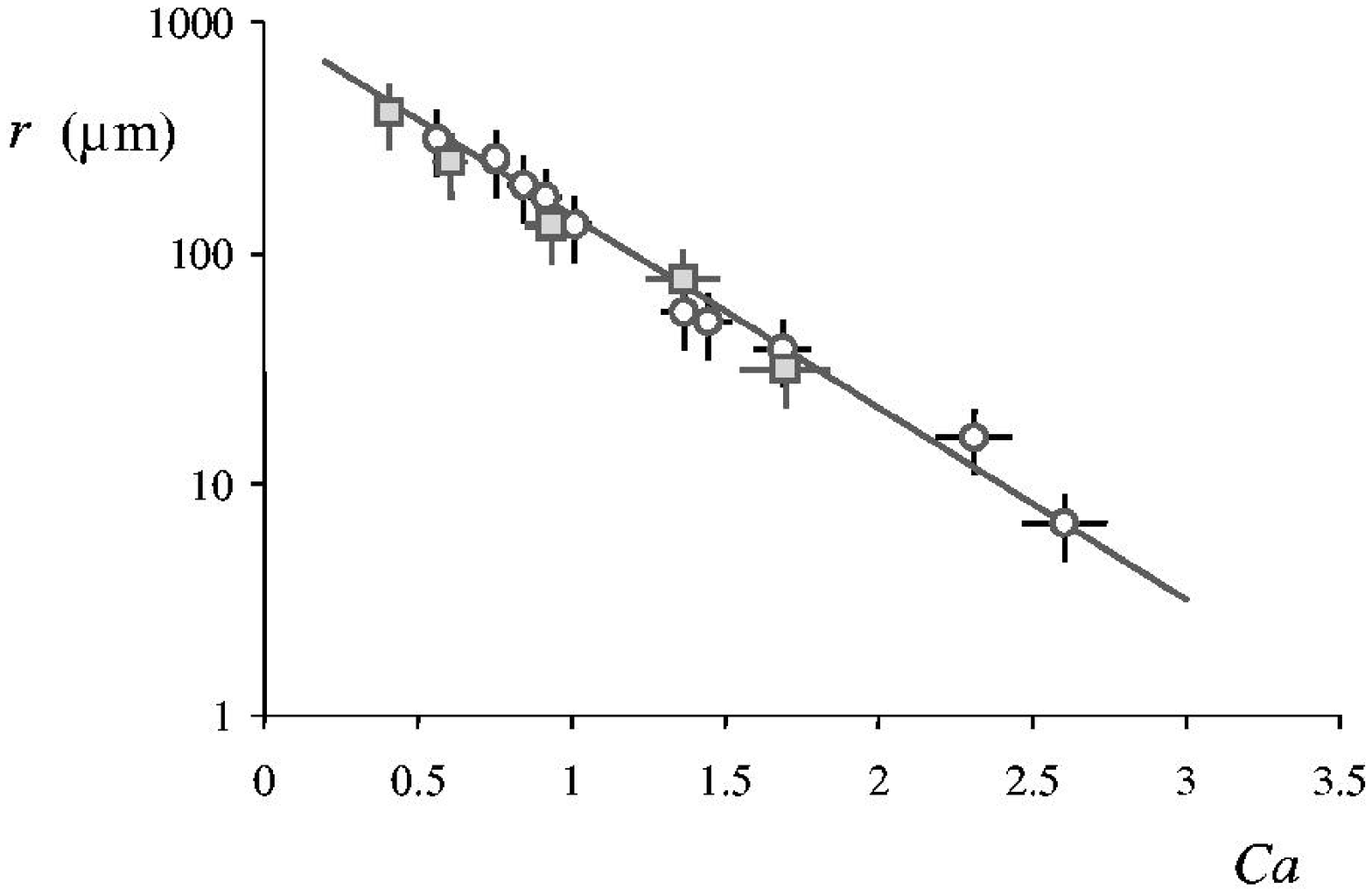}
\caption{
Experimental data on the cusping of a viscous fluid, taken from \cite{LRQ03}.
On the left, a closeup of the tip of a cusp on the surface of a viscous 
fluid; the scale bar corresponds to 200 $\mu$ m. On the right, the radius 
of curvature of the (almost) cusp as function of capillary number. 
In agreement with (\ref{radius_asymp}), the dependence is exponential. 
    }
\label{cusp_fig_exp}
\end{figure}
It is easy to confirm that (\ref{surf_JM}) yields a cusp for $a = -1/3$,
i.e. for $Ca = \infty$ or vanishing surface tension. If one expands 
around the cusp point by putting $\theta=\pi/2 + \delta$, one obtains
\begin{eqnarray}
\label{cusp_JM}
&& x = -\frac{2\epsilon}{3}\delta - \frac{\delta^3}{12} \\
&& y + \frac{2}{3} - 2\epsilon = \frac{\delta^2}{6},
\end{eqnarray}
which is the generic cusp scenario (\ref{cusp}). As is apparent
from Fig.~\ref{cusp_fig}, the case $\epsilon<0$ leads to self-intersection
of the free surface, which is of course not physical. The radius 
of curvature of the cusp for $\epsilon>0$ is given by 
\begin{equation}
R \approx \frac{256}{3} \exp\left\{-32\pi Ca\right\}, 
\label{radius_asymp}
\end{equation}
as found from (\ref{cusp_JM}),(\ref{a_asymp}). The exponential 
dependence (\ref{radius_asymp}) has been confirmed experimentally
(cf. Fig.~\ref{cusp_fig_exp}).

\section{Born-Infeld equation}
\label{sec:Born}
The ideas presented here are not restricted to free surface problems. 
As an illustration we present a problem that appears in connection with 
string theory \cite{Tsey99}, but is also of long-standing interest in
the theory of non-linear waves \cite{Witham}. 
The Born-Infeld equation reads
\begin{equation}
z_{tt}(1+z_x^2) - z_{xx}(1-z_t^2) = 2z_x z_t z_{xt}
\label{BI}
\end{equation}
Hoppe \cite{Ho95} gave a general solution of the form
\begin{eqnarray}
\label{Ho1}
x'(t,\varphi) = \lambda\cos(f-g)\cos(f+g),\quad
z'(t,\varphi) = \lambda\cos(f-g)\sin(f+g) \\
\label{Ho2}
\dot{x}(t,\varphi) = -\sin(f-g)\sin(f+g),\quad
\dot{z}(t,\varphi) = \sin(f-g)\cos(f+g),
\end{eqnarray}
where $f=f(\varphi+t/\lambda)$ and $g=g(\varphi-t/\lambda)$ are any
two (smooth) functions and $\varphi$ is a parameter. For (\ref{Ho1})
to be a graph, the tangent vector (\ref{Ho1}) should not be vertical,
i.e. we must require that $|f + g| < \pi/2$.

The curvature of this solution is
\begin{equation}
\kappa(t,\varphi) = \frac{f'+g'}{\cos(f-g)},
\label{curv}
\end{equation}
so a singularity is expected whenever $f-g = \pi/2$. Let us assume for
simplicity that $g(\zeta) = - f(-\zeta)$. I don't expect this to be
a restriction on the class of possible singularities that can occur.
Let $f(\zeta)$ have the local expansion
\begin{equation}
f(\zeta) = \pi/4 + a(\zeta-\zeta_0) - b(\zeta-\zeta_0)^2 +
O(\zeta-\zeta_0)^3,
\label{f_exp}
\end{equation}
so together with the symmetry requirement we find
\begin{equation}
f - g = \pi/2 + 2a(t-\zeta_0) - 2b(\varphi^2+(t-\zeta_0)^2).
\label{f_minus}
\end{equation}
The factor $\lambda$ can be absorbed into $a$.

From this expression it is clear that $\zeta_0$ has to be identified
with the singular time $t_0$ and $a>0$ for the solution to be regular
for $t<t_0$. To expand around the singular time, we put $t' = t_0 - t$.
Similarly, one must have $b>0$ (otherwise $f-g$ would be $\pi/2$ at an
earlier time), and the singularity occurs at $\phi = 0$. Thus to
leading order we have
\begin{equation}
f - g \approx \pi/2 - 2a t' - 2b\varphi^2, \quad
f + g \approx 2a\varphi,
\label{f_sum}
\end{equation}
from which we get
\begin{equation}
x' = 2at' + 2b\varphi^2, \quad z' = 4a^2t'\varphi + 4ab\varphi^3.
\label{xp}
\end{equation}
Integrating this expression, using the integrability condition
(\ref{Ho2}), gives
\begin{equation}
x = t'\varphi + 2c\varphi^3/3, \quad z = t'\varphi^2/2 + c\varphi^4,
\label{BI_swallow}
\end{equation}
where we have used a rescaling of the parameter $\varphi$. This is
of course exactly the swallowtail (\ref{swallow}) with $\gamma = 1$.

\end{document}